# Optimization of CMOS image sensors with single photon-trapping hole per pixel for enhanced sensitivity in near-infrared

E. Ponizovskaya Devine[1,2], Ahasan Ahamed[2], Ahmed S. Mayet[2], Soroush Ghandiparsi[2],
Cesar Bartolo-Perez[2], Lisa McPhillips[2], Aly F. Elrefaie[1], Toshishige Yamada[1,3], Shih-Yuan Wang[1], M Saif Islam[2]

[1] *W&WSens Device, 4546 El Camino, Los Altos, CA 94022 USA*
[2] *University of California, Davis, CA 95616 USA*
[3] *University of California, Santa Cruz, Santa Cruz, CA 95064 USA*

**Abstract—** The optimization of silicon photodiode-based CMOS sensors with backside-illumination for 300–1000 nm wavelength range was studied. It was demonstrated that a single hole on a photodiode increases the optical efficiency of the pixel in near-infrared wavelengths. A hole with optimal dimensions enhanced optical absorption by 60% for a 3 µm thick Si photodiode, which is 4 orders better than that for comparable flat photodiodes. We have shown that there is an optimal size and depth of the hole that exhibits maximal absorption in blue, green, red, and infrared. Crosstalk was successfully reduced by employing thin trenches between pixels of 1.12 µm$^2$ in size.

## I. INTRODUCTION

The growing demand for mobile imaging, digital cameras, surveillance, monitoring, and biometric evaluation has promoted interest in complementary metal-oxide-semiconductor (CMOS) image sensors [1–3]. These applications demand high sensitivity and resolution in the near-infrared. Recent approaches to increase pixel density typically involves reducing the pixel pitch from 2.2 µm to less than 0.8 µm [3, 4]. Such an approach suffers from a loss of optical sensitivity due to the reduction of the absorption area. A thicker absorption layer can be used to offset this effect, but this also reduces the speed of operation—and in turn, the bandwidth of the imager. Smaller pixel sizes may also have a parasitic charge exchange between neighboring pixels (crosstalk) [5]. Microholes are known to increase the crosstalk between pixels but this problem can be limited with deep trench isolation (DTI) [6–8].

Recently, it has been proposed to implement an array of nanoholes [9, 10], or even a single nanohole [6] on a pixel's surface to mitigate the trade-off between resolution, efficiency, and speed. Two different methods are implemented for near-infrared imaging: (i) application of a separate near-infrared filter, based on plasmonic techniques [11], or gratings [12]; and (ii) standard low-cost pigment filters [13] that are transparent in near-infrared wavelengths.

In this work, we optimize the implementation of a single microhole on a pixel by varying the size and depth of either a cylindrical or inverted pyramid hole to enhance the optical efficiency of the imagers in the near-infrared. The enhanced optical efficiency in the near-infrared spectrum was compared to the same structure without the microhole.

For this study, we consider the stacked low-noise structure similar to the one proposed by Sony Corporation [14], which is back-side illuminated and layered over a chip with signal processing circuits. Besides just improving the chip size, stacking technology also improves the dark characteristics (such as noise and white pixels), by optimizing the sensor process independently.

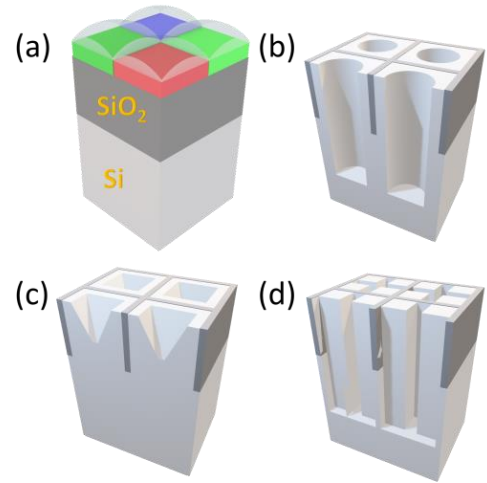

Fig. 1. Schematic diagram of CMOS image pixels: View of the pixels with micro lenses, color filters, microholes and Deep Trench Isolation (DTI). (a); Cross-section view of the image pixels (without the micro lens and color filter) showing inverted pyramid holes (b), cylindrical holes (c), and cross trenches (d). DTI separation between the pixels is also shown.

To increase the quantum efficiency, the surface of each pixel is etched with a small, inverted pyramid with dimensions in the range of 400 nm [3] and we model a Bayer array with a standard color filter

Contact author: E. Ponizovskaya Devine (eponizovskyadevine@ucdavis.edu)

array [15] [Fig. 1 (a)-(d)]. Microhole arrays are known to dramatically increase the optical efficiency of Si in the 800-1000 nm wavelength range [9]. Here, we use a single hole per diode instead. The hole size is in the range of 800-1000 nm. We simulated CMOS image sensors with different hole shapes, such as an inverted pyramid, cylindrical, and crossed rectangular holes [6], and we compared the optical efficiency and the crosstalk for each shape. The deep trench structure [7, 8] with Si-SiO2 interface is used as a barrier against electron diffusion and assists in confining light within the pixels by acting as a reflector.

We simulated 1.12 µm wide and 3 µm deep pixels with trenches of 150 nm width and 2.5 µm depth that showed a reasonable reduction in crosstalk [6]. Previous simulations [6] also showed that a single funnel hole in a device provides better efficiency compared to an array and reaches up to 70% optical efficiency. We varied the dimensions of the cylindrical holes, inverted pyramid holes, and crossed rectangular holes to determine their optimal parameters. As was previously shown, the holes with a size comparable to the wavelength can change the light's direction of propagation and bend it laterally, which provides a better light trapping effect and reduces reflection [16–18]. This technique is implemented here for high-speed Si photodetectors. We used the Lumerical FDTD software, and the simulation methodology described in [1] to calculate the absorption of each pixel. The optimal single hole enhanced the absorption at 850 nm and 940 nm wavelength to more than 60% for a pixel of 1.12×1.12 µm lateral dimension and 3 µm depth. This device is two times thinner than the devices reported in [1].

## II. OPTICAL SIMULATION METHODOLOGY

The microstructure is capable of redirecting the normal incident light into the lateral directions parallel to the plane of the surface. Light absorption is higher in the lateral direction while the generated electro-hole pairs are collected through the intrinsic Si layer due to the modes that stay in the structure until they are absorbed [9]. We have shown the effectiveness of this method before [6] in the CMOS pixels. Moreover, it was shown that even one hole of the right size can significantly increase a pixel's optical efficiency.

The Finite Difference Time Domain (FDTD) [19] provided by the Lumerical software package [20] was used to solve Maxwell's curl equations numerically for the unit cell of the Bayer array (Fig. 1). We considered several hole shapes, such as the cylindrical, the inverted pyramids, and the crossed rectangular holes [6]. Bloch boundary conditions in the XY plane and Perfectly Matched Layer (PML) in the direction normal (Z) to the surface were also considered. The CMOS image sensor model includes red, green, and blue filters with a thickness of 900 nm, lenses, antireflection coating, and a 3 µm thick Si on SOI substrate, and micro-lens with a radius of 1 µm and a thickness of 500 nm. The wavelength of the normal plane wave source varied in the range of 300 and 1000 nm. The methodology was described in [6].

The optical efficiency (OE) is given by the Poynting vector normal to the surface (P) measured around the cells:

$$OE = \frac{P_{in} - P_{out}}{P_{inc}} \quad (1)$$

Where $P_{inc}$ is the Poynting vector of the incident light calculated above the lenses and filters. We assume that the quantum efficiency is proportional to the optical efficiency [21]. The filter's response spectrum is presented in Fig. 1a by the dashed line and the real and imaginary parts of the refractive index n and k are shown in the appendix. For the simulation setups, the transmittance of the pigment filters with 900 nm thickness is used [13]. The transmission profile of the flat surface device (with no photon-trapping structure on the surface) is shown in Fig. 2b as a reference. The maximum optical efficiency for the Bayer filter is determined by its transmittance (Fig. 2a). For these simulations, they are $OE_{blue}$ = 80% at 440 nm, $OE_{green}$ = 65% at 550 nm, and $OE_{red}$ = 85% at 650 nm, and transparent in the near-infrared. The Si optical absorption is very weak in the near-infrared spectrum (close to its bandgap 1.12eV) and light-trapping strategies are required to enhance its absorption for higher optical efficiency.

## III. RESULTS AND DISCUSSIONS

Fig. 2a shows that transmittance through the filter limits the maximum absorption. In Fig. 2b shows the optical efficiency (OE) for the flat sensors without DTI, as determined by the FDTD optical simulations. Here, we observe the OE in the infrared wavelength to be very low as expected in contrast to the nominal OE for the visible wavelengths in blue, red, and green spectra. There is relatively lower crosstalk for vertical illumination even without trenches.

We optimized the parameters of the holes by varying their size and depth. We choose the diameter of the cylindrical holes to be 700nm since it was demonstrated to be optimal in [6, 9, 10] and varied their depths. Pyramid dimensions are correlated due to the fabrication process, so only the length of the sides is varied. The crossed rectangular hole is composed of two, overlapping rectangular holes that are crossed in the center and positioned to reach the trenches on each side. We vary the width and the depth of the rectangular holes. The optimization process can be understood from the results shown in Fig. 3 for the 850 and 940 nm wavelengths. All pixels, blue, green, and red have the same OE in infrared. The results in Fig. 3a show the OE for different hole depths for the cylindrical holes with diameters of 700 nm. We find the optimal depth for the cylindrical holes to be about 2000 nm. The variation of the depth shows that OE in infrared increases with the depth and stabilizes after 2000 nm. The OE in blue also increases up to a depth of 2000 nm but begins to decrease after that. This could be explained by how the maximum absorption corresponds to a resonant behavior that happens at the smaller size for the smaller wavelength. A similar variation on the inverted pyramid side length (Fig. 3b) indicates that the optimal side length is close to 700-800 nm for each pyramid edge. The pyramid size optimizations showed that increasing size produces better absorption for the near-infrared and reduces the absorption for the blue pixels. The optimal side length of 700 nm results in a balance between the OE of the blue and the near-infrared wavelengths. The crossed rectangular holes' depth and width were also optimized. Fig. 3c shows that the optimal depth for the crossed rectangular holes is about 2200-2500 nm, with a width of 250 nm. It is about the same as for cylindrical holes. The impact of the rectangular holes' width on OE is almost the same over the range from 150-400 nm (Fig. 3d), the optimal width is about 200-300 nm for the depth of 2500 nm. The OE of such image sensors calculated for a longer wavelength of 940 nm are encouraging and reach over 50%, while for wavelengths longer than 1 µm, the OE rapidly decreases. Microholes provide smaller reflection for reasons similar to the Lambertian reflector [22]. It assists in trapping light within the Si and bending the normal incident light into lateral modes, as was numerically and experimentally shown in [10, 23]. While a guided resonant mode can be completely absorbed, high absorption can still be achieved with leaky modes that can propagate in the device long enough to be mostly absorbed. The best results for the cylindrical holes, inverted pyramids, and crossed rectangular holes are shown in Fig. 4. All the microstructures increase the absorption in the pixels for blue, green, and red with higher enhancement in the green and red wavelength spectrum.

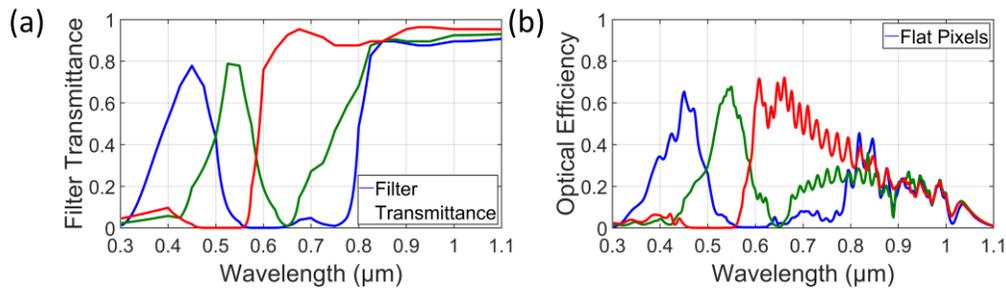

Fig. 2. Simulated transmittance after filters (a), and optical efficiency (OE) of a flat pixel without DTI (b).

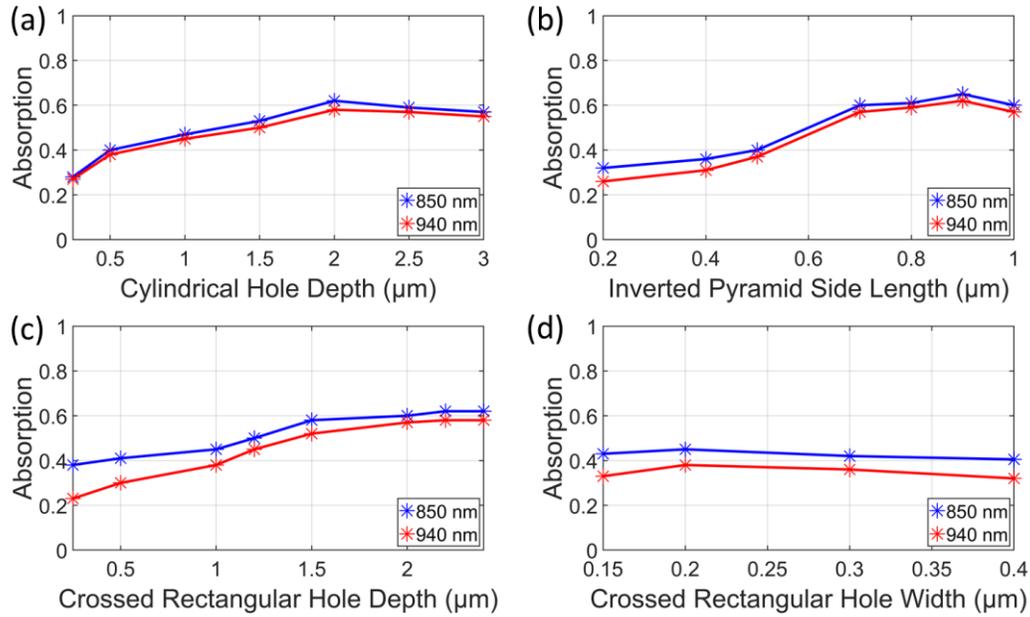

Fig. 3. Optimization: optical efficiency (OE) vs cylinder depth for cylindrical holes (a), inverted pyramid side length (b), crossed rectangular hole depth (c), and crossed rectangular hole width (d).

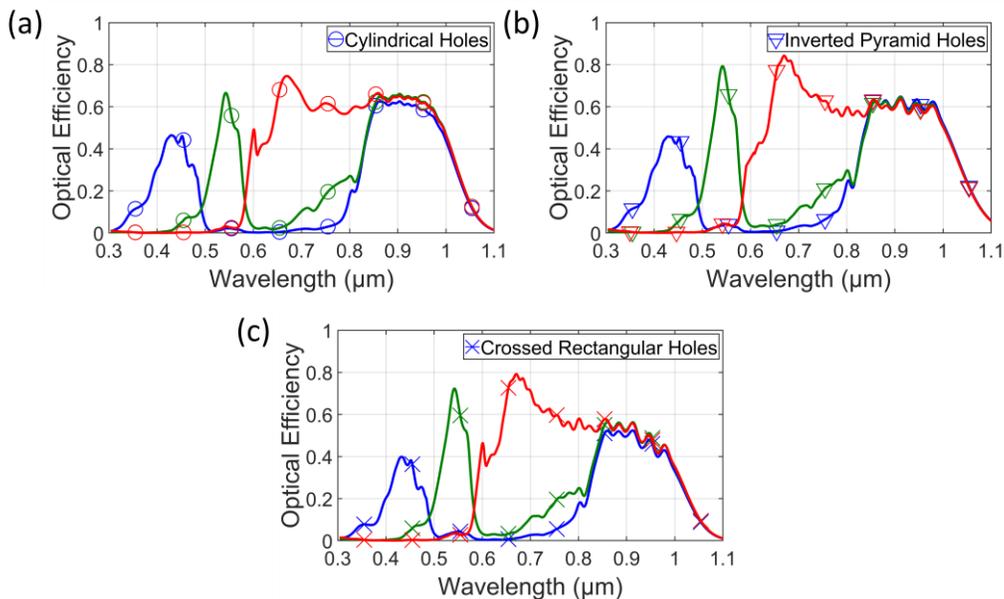

Fig. 4. OE of a single pixel with optimized dimensions of cylindrical holes (a), inverted pyramid holes (b), and crossed rectangular holes (c) respectively with DTI.

Despite enhancing the OE, the integration of holes in the devices can increase the crosstalk between pixels. As mentioned, the crosstalk in the visible region is smaller than the near-infrared owing to the inherent material properties of the crystalline silicon. Hence, the color separation in the blue and green region is better than the green and red region, as it is expected to obtain a better color separation and low color error there [21]. However, as we can see from Fig. 4 the crosstalk was effectively reduced by the implementation of trenches without any decrease in the OE compared to that exhibited by the flat pixels.

## IV. CONCLUSION

We have optimized the optical efficiency (OE) of silicon CMOS image sensors with backside-illumination designed with 3 µm thick silicon and with 1.12 µm$^2$ pixel size in RGB. Optical FDTD simulations have shown the optimal dimensions of a single-hole pixel in the shape of cylinder, inverted pyramid, or a crossed rectangular hole. In the near-infrared wavelengths, the OE of the sensors for the tested geometries is slightly smaller than that of the funnel holes [6]. However, it still attains values higher than 60%. The optimized structures, diameters, and depths for a single-hole pixel provides insight for designing enhanced, optically sensitive image sensors in near-infrared. Our simulation shows that the increased optical absorption is caused by the photon-trapping hole, which supports lateral modes and confines the light within the pixel. Different hole shapes and depths were investigated and optimized based on the modal analysis. A single-hole pixel has increased crosstalk; however, implementing trenches with dimensions of 250 nm width and 2.5 µm depth between pixels efficiently reduces the crosstalk to the normal level.

TABLE I: FILTERS PARAMETERS

| Wavelength, nm | Blue n | k | Green n | k | Red n | k |
|---|---|---|---|---|---|---|
| 300 | 1.55 | 0.055 | 1.62 | 0.4 | 1.54 | 0.415 |
| 400 | 1.54 | 0.07 | 1.6 | 0.5 | 1.54 | 0.325 |
| 500 | 1.54 | 0.215 | 1.58 | 0.405 | 1.53 | 0.155 |
| 600 | 1.54 | 0.8 | 1.57 | 0.05 | 1.53 | 0.048 |
| 700 | 1.53 | 0.45 | 1.57 | 0.11 | 1.52 | 0.033 |
| 800 | 1.53 | 0.455 | 1.57 | 0.105 | 1.52 | 0.375 |
| 900 | 1.52 | 0.465 | 1.56 | 0.09 | 1.52 | 0.0185 |
| 1000 | 1.52 | 0.39 | 1.56 | 0.055 | 1.52 | 0.015 |

## ACKNOWLEDGMENT

We acknowledge W&WSens Devices for financial support.

## APPENDIX

The complex optical constants (n+ik) of color filters, where the real part of the optical constants is called refractive index (n) and the imaginary part is called the extinction coefficient (k), was used in the optical simulation of the image sensors. The data was taken from [7], [8]. Table I represents the parameters of the filters which were used in the simulations.